\documentclass[twocolumn,aps,prl]{revtex4}

\usepackage{graphicx}
\usepackage{dcolumn}
\usepackage{bm}

\begin{document}


\title{Laser beam filamentation in fractal aggregates}
\author{Claudio Conti$^{1,2}$, Neda Ghofraniha$^3$,  Giancarlo Ruocco,$^{2}$ Stefano Trillo$^{4}$}
\email{claudio.conti@phys.uniroma1.it}
\affiliation{
$^{1}$Research Center ``Enrico Fermi'', Via Panisperna 89/A, 
00184 Rome, Italy \\
$^{2}$INFM CRS-SOFT Universita' di Roma ``La Sapienza'',  P. A. Moro 
2, 00185, Roma, Italy\\
$^{3}$INFM SMC Universita' di Roma ``La Sapienza'',  P. A. Moro 
2, 00185, Roma, Italy\\
$^{4}$Dept.~of Engineering, University of Ferrara, Via Saragat 1, 
44100 Ferrara, Italy
}
\date{\today}  
\begin{abstract}
We investigate filamentation of a cw laser beam in soft matter
such as  colloidal suspensions and fractal gels. The process,
driven by electrostriction, is strongly affected by material properties,
which are taken into account via the static structure factor,
and have impact on the statistics of the light filaments.
\end{abstract}

\pacs{42.65.Jx, 42.65.Tg, 82.70.-y}
\maketitle 
Light propagation in soft-matter such as colloidal systems can be affected via a feedback
mechanism by light-induced structural changes.
In spite of the low powers needed to observe such phenomena
\cite{Ashkin82}, the optical nonlinear response of soft matter was generally overlooked
with the exception of metal colloids \cite{polariton}
and liquid crystals \cite{Khoo95}.
\newline\indent
Here we consider the case of a colloidal suspension of 
dielectric particles dispersed in a solvent, characterized
by tunable interactions that can be responsible for long range 
correlations and even phase transitions such as gelation.
The role of the structure of these materials in nonlinear optical processes 
is essentially unexplored. Large scale ordered structures can affect 
orientational \cite{Khoo95}, electroctrictive \cite{Ashkin82,Boydbook}, and thermoforetic \cite{Piazza02} mechanisms.
For isotropic particles and negligible thermal gradients due to light absorption,
the leading mechanism is expected to be electrostriction: 
the particles are subject to forces induced by light intensity gradients, 
thus moving in the region with higher or lower intensity
depending on the difference between their refractive index $n_s$ and that of the host medium $n_h$ \cite{Boydbook}.
In both cases, the optical beam experiences self-focusing, a process which has been described by a local Kerr law or
intensity-dependent refractive index variation $\Delta n=n_2 I$,
where $n_2$ is the Kerr coefficient and  $I$ the optical intensity \cite{Ashkin82}. 
Recently, it has been shown that the nonlinear response of such materials
is mediated by the static structure factor $S(q)$ of the material \cite{SimpleLiquids}, 
thus turning out to be strongly nonlocal \cite{conti05prl}.
While the strength of the nonlinearity ($n_2$) is given by the material 
compressibility, $S(q)$ is strongly affected by the whole structure of the soft-material phase, 
e.g. by the presence of fractal aggregates which has impact on the nonlinear susceptibilities.
\newline\indent
In this Letter, we investigate laser beam filamentation, 
a process which is well known to occur due to spatial modulational instability (MI) \cite{bespalov66}.
MI is ubiquitous in nonlinear science, occurring in fluidodynamics
(Benjamin-Feir instability  \cite{benjamin67b}),
matter waves \cite{saffman98},
plasma physics\cite{taniuti68},
coherent oscillations in lattices \cite{leon99},   
and spin waves \cite{lai98}.
In the context of optics, filamentation has been recently shown to be affected by the coherence property of light \cite{kip00},
by the artificial periodicity of nonlinear media \cite{corney01}, and by nonlocality \cite{wyller02}.
Specifically, filamentation has been investigated in liquid crystals
where the reorientational nonlinearity allows to tailor the nonlocality \cite{peccianti03}, 
and is at the origin of complex light dynamics \cite{conti05PRE}. 
It is therefore interesting to investigate how the dominant nonlinear mechanisms
in soft matter, and particularly in complex media such as fractal aggregates,
affect the filamentation process.
In addition, the considered light induced perturbations can be reversible or irreversible (depending on the specific material
relaxation dynamics); in the latter case the filaments can be used for the all-optical imprinting of
optics circuitry.

We start from the following model for the electrostrictive response
of soft matter \cite{conti05prl}
\begin{equation} \label{unidir}
i \frac{\partial E}{\partial z} + \frac{1}{2k}\nabla_{\perp}^2 E +
k_0 \left(\frac{\partial n}{\partial \rho}\right)_{\rho_0} \rho E=0\text{,}
\end{equation}
where $E$ is the complex envelope of the electric field at angular frequency 
$\omega$, $k=k_{0} n_0=(\omega/c) n_0$ is the wavenumber, 
$n_0$ is the average refractive index, 
and $\nabla^2_{\perp}=\partial_x^2+\partial_y^2$.
Here $\rho$ is the optically induced perturbation to the bulk particle density $\rho_0$, 
which is related to the optical intensity through the static structure factor \cite{conti05prl}:
$\tilde \rho(q)=\gamma_e Z_0 S(q) \tilde I(q)/2 n_0 k_B T$,
where tilde denotes Fourier transform, $\gamma_e=\rho_0 (\partial \epsilon/\partial \rho)_{\rho_0}$ 
the electrostrictive coefficient \cite{Boydbook} 
$Z_0$ is the vacuum impedance, $k_B$ is the Boltzmann
constant, and $T$ is the system temperature.

\begin{figure}
\includegraphics[width=8.3cm]{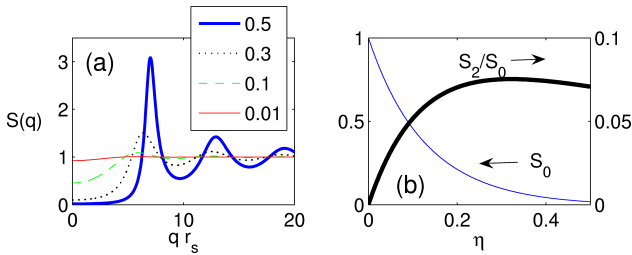}
\caption{(Color online) Properties of HS colloids:
(a) $S(q)$ vs. normalized wavenumber $q r_{s}$ 
for different values of packing fraction $\eta$; 
(b) $S_0$ (nonlinearity) and $S_2/S_{0}$ (nonlocality) vs. $\eta$.
\label{figurehs}}
\end{figure}

MI is the process that triggers the generation of multiple filaments
due to exponential amplification of plane wave perturbations with small
transverse wavenumbers $q^2=q_x^2+q_y^2$
(changes in longitudinal wavenumber $q_z$ of the perturbation is commonly
assumed to be negligible due to the paraxial conditions), 
at the expenses of a pump beam $E=E_0\exp(i \beta z)$.
A linear stability  analysis \cite{peccianti03,wyller02}  allows us to find the growth rate or MI gain $G(q_x, q_y)$ as
\begin{equation} \label{gain}
G z_d=q w_0 \sqrt{\frac{P}{P_r}\frac{S(q)}{S_0}-(q w_0)^2}
\end{equation}
where $w_0$ is a reference waist (of intensity), 
$z_d=k w_0^2$ is the associated diffraction (Rayleigh) length, 
$P=\pi w_0^2 I_0$ is the beam power, with $I_0=n_0 E_0^2/(2Z_0)$ the peak intensity,
and $P_r \displaystyle \equiv \pi k_B T \rho_0 \epsilon_0^2 c n_0 /(S_0 \gamma_e^2 k_0^2)$ 
is a reference power ($P_r \cong 3 \mu W$ for typical numbers $r_s=10$ nm, $w_0=100$ $\mu$m,
$n_s=1.5$, $n_h=1.3$, and a particle density  $\rho_0=10^{21}$ m$^{-3}$).
The Kerr limit is recovered in Eq.~(\ref{gain}) for a constant $S(q)=S_0$,
where $S_0=S(0)$ is the ratio between the material and the ideal gas compressibility, and is proportional to the Kerr coefficient 
$n_2=Z_0 \gamma_e^2 S_0 / 4\epsilon_0 n_0^2 k_B T \rho_0$.
Let us consider, first, the case of colloids made by hard spheres (HS)
with radius $r_s$ and dielectric constant $\epsilon_s$, for which the electrostrictive 
coefficient is $\gamma_e=4\pi r_s^3 \epsilon_h  (\epsilon_s-\epsilon_h)/(\epsilon_s+2\epsilon_h)$,
$\epsilon_h$ being the dielectric constant of the host medium (e.g. water).
In this case, $S(q)$, from \cite{HS},
is displayed in Fig.~\ref{figurehs}a for different values of packing fraction $\eta$ of the spheres. 
At very low packing fractions the medium behaves as a Kerr medium with an approximately constant $S(q)$. 
However, the curvature of $S(q)$ at low wave-numbers increases with $\eta$, until a clear peak appears
for  tightly packed spheres. This means that material response becomes more nonlocal as 
spheres are more densely packed. At the same time, however, the nonlinearity (compressibility)
decreases.
A parabolic approximation $S(q) \cong S_0+S_2 (q r_s)^2=S_0+K q^2$ 
allows us to show explicitly the dependence of nonlinearity ($S_0$) and nonlocality ($S_2/S_0$)
on $\eta$ (see Fig. \ref{figurehs}b). However, we argue that nonlocal behavior 
can be hardly seen for HS. In fact, by defining a {\em degree of nonlocality}  \cite{wyller02} 
$\sigma^2 \equiv K/(S_0 w_0^2)=(S_2/S_0)(r_s/w_0)^2$, 
for typical number, e.g. $w_0=100$ $\mu$m  and particle size $r_s=10$ nm, 
we find $\sigma^2 \cong 10^{-9}$ [$S_2/S_0 \cong 0.1$, see Fig. \ref{figurehs}b].
Hence for HS a local model works quite well \cite{Ashkin82}, even at high packing fractions, 
as expected on the basis of the extremely short interaction range involved.

\begin{figure}
\includegraphics[width=8.3cm]{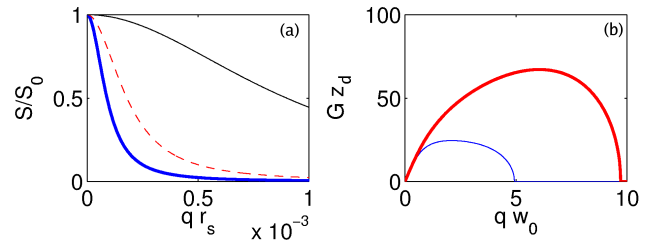}
\caption{
(Color online) (a) $S(q)$ in the presence of aggregration, 
for $D=2.2$ and $\xi=10^4 r_s$ (thick line),
$\xi=5\times10^3 r_s$ (dashed line) and $\xi=10^3 r_s$ (thin line); (b) MI gain when $P=10^3 P_r$ and $\xi=w_0=10^4 r_s$,
with $D=2.2$ (thin line) and $D=1.2$ (thick line).
\label{Sqcluster}}
\end{figure}

The situation becomes drastically different when long range correlation sets in 
due to induced particle aggregation, as for instance in gelly or glassy arrested states.  
Either kinetic or structural arrest results in a peak in $S(q)$ at $q=0$  
due to enhanced compressibility and long range spatial correlations.
One possible way to realize kinetic, out of equilibrium, gelation process is 
diffusion limited cluster aggregation (DLCA)
induced by a strong and very short-range attraction  
between particles in addition to HS repulsion \cite{DLCA}. 
This means that once two particles get close
they stick irreversibily resulting into a so called 
fractal gel, i. e. a space filling network of interconnected fractal clusters,
characterized by well known expressions of the structure factor 
(see e.g. Eq. 4 in \cite{selfsimilar}).
Such $S(q)$, parametrized by the long range correlation 
distance (aggregate dimension) $\xi$ and by the fractal dimension of the aggregates $D$,
is reported in Fig. \ref{Sqcluster}a.
The corresponding gain profile [from Eq. (\ref{gain})] exhibits
a single band as shown in Fig. \ref{Sqcluster}b.
\begin{figure}
\includegraphics[width=8.3cm]{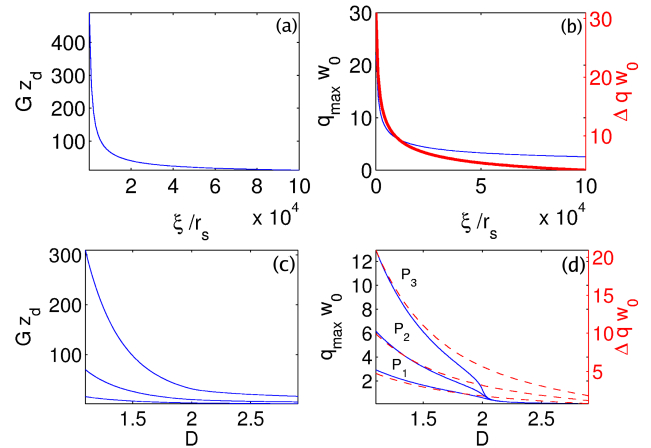}
\caption{
(Color online) Dependence of MI features on aggregate dimension $\xi$:
peak gain $G z_d$ (a) and maximally amplified wavenumber (blue line, left ordinates)  $q_{max}$ 
and bandwidth  (red line, right ordinates) $\Delta q$ (b) vs. $\xi/r_{s}$. 
Here $P=10^3 P_r$, $D=1.5$, $w_0=10^3 r_s$.
(c-d) shows the same quantities as a function of fractal dimension $D$ for
fixed $\xi=10 w_0$ ($w_0=10^4 r_s$), and powers $P_1=10^3 P_r, P_2=10^4 P_r, P_3=10^5 P_r$.
\label{MIclusterVsxi}}
\end{figure}
\begin{figure}
\includegraphics[width=8.3cm]{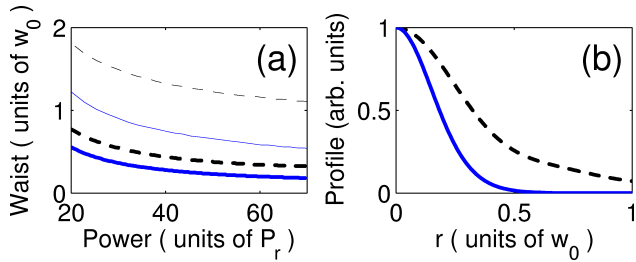}
\caption{(Color online) (a) Waist (FWHM) of the solitary wave solutions (solid lines) 
and associated index perturbation (dashed lines) vs. their power for $D=1.3$ (thick) and $D=2.3$ (thin).
Here $w_0=10^4 r_s$ and $\xi=10 w_0$.
(b) Intensity and index profile sampled at $P=70 P_r$ ($D=1.3$).\\
\label{indexwaist}}
\end{figure}
The fractal dimension $D$ and the aggregates characteristic length $\xi$ 
have a strong impact on the characteristics of MI.
First, comparing Figs. \ref{Sqcluster}a and \ref{figurehs}a,
it is clear that $S(q)$ varies on a $q-$scale much smaller
and the nonlocality 
is much higher in the presence of aggregates.

Specifically, from the expansion $S\cong S_0-Kq^2$ 
we find $\sigma^2 \equiv K/(S_0 w_0^2)=D(D+1)(\xi/w_0)^2/6$, 
showing that (for a fixed fractal dimension $D$) the nonlocality grows
with the relative (to beam waist) cluster dimension $\xi/w_0$. 
Conversely if the beam waist contains many aggregates instead of a spanning network,
the situation becomes similar to that of many independent particles in the beam, 
and hence the system behaves locally.

The dependence of the MI feature on the aggregate dimension are summarized 
in Fig. \ref{MIclusterVsxi} for a typical value of power $P=10^3 P_r$ (no qualitative change 
are observed by varying $P$).
Importantly, in spite of the fact that $n_2$ increases with $\xi$ 
($S_0$ grows due to the higher compressibility), the enhanced nonlocality
acts as a filter that frustrates MI, and cause the peak gain $G$, the MI bandwidth $\Delta q$ or 
the maximally amplified spatial wavenumber $q_{max}$ to decrease dramatically with $\xi$. 
This is relevant when the colloidal sample undergoes phase transitions such as gelation,
because the filamentation pattern is expected to change as the long range correlation 
associated with material structural changes developes.
The dependence of the MI 
gain on the fractal dimension $D$ is summarized
in Fig. \ref{MIclusterVsxi}c,d.
Notably for $D<2$ the MI process appears more pronounced and $q_{max}$ strongly varies with $D$.
We emphasize, however, that smaller $D$ implies smaller $S_0$, 
and hence larger powers $P_r$.

Beyond the early stage of exponential amplification, MI is expected to give rise to filaments
corresponding to solitary waves of Eq. (1) \cite{conti05prl}. In this sense, the nonlocality plays a key role,
acting as a spatial filter that tends to flatten the induced index perturbation, 
thus stabilizing the single filaments. This is shown in  Fig. \ref{indexwaist}, where we report
the FWHM (full-width-half-maximum, henceforth waist) of the intensity and associated index change profiles 
against filament power (the profiles are shown in Fig. \ref{indexwaist}b).
Unlike local Kerr media, the index perturbation associated with the solitary waves
turns out to be broader than the intensity distribution, and its waist tends to saturate.
Once again, the fractal dimension plays a crucial role determining the size of filaments.

In order to provide more direct evidence that the output filament pattern
is strongly affected by the fractal dimension $D$, 
we resorted to the numerical integration of Eqs. (\ref{unidir}). 
The key point is to recognize that in soft matter, the leading mechanism that triggers laser beam filamentation, 
is the presence of material density fluctuations. 
These are known to be ruled by $S(q)$,
and, in the spectral domain, can be written as $ \tilde{\rho}=S(q) \tilde{f}$, 
where $f$ is a 3D stochastic variable (white noise).
To model this effect we generalize Eq. (\ref{unidir}) by resorting to the following 
stochastic PDE model 
\begin{eqnarray}
\label{unidirnorm}
i\frac{\partial a}{\partial \zeta}+\frac{1}{2} \nabla^2_{st} a+\psi a=0\,;\;
\tilde \psi=R(q_s,q_t) (\widetilde{|a|^2}+\tilde \nu),
\end{eqnarray}
where we have introduced the dimensionless quantities 
$R=S/S_0$, $(s,t)=(x,y)/w_0$, $\zeta=z/z_d$, 
and $\tilde \nu$ is a spectral noise term such that $\langle \nu(s,t,\zeta) \nu(s',t',\zeta') \rangle=\nu_p^2 \delta(s-s') \delta(t-t') 
\delta(\zeta-\zeta')$, the brackets denoting statistical average.
The density is given by $\rho= \rho_N \psi$, with $\rho_N=[k_0 z_d (\partial n/\partial \rho)]^{-1}$,
and $\psi$ is the sum of a nonlinear term and of a colored (through $R=S/S_0$) noise.
Equations (\ref{unidirnorm}) are solved by the Heun method \cite{stochPDE}, 
after a FFT pseudospectral discretization along $s$ and $t$.
We show results obtained with a plane wave input
$a(\zeta=0)=a_0$ ($a_0^2=P/4 P_r$ due to simple scaling arguments).
\begin{figure}
\includegraphics[width=9cm]{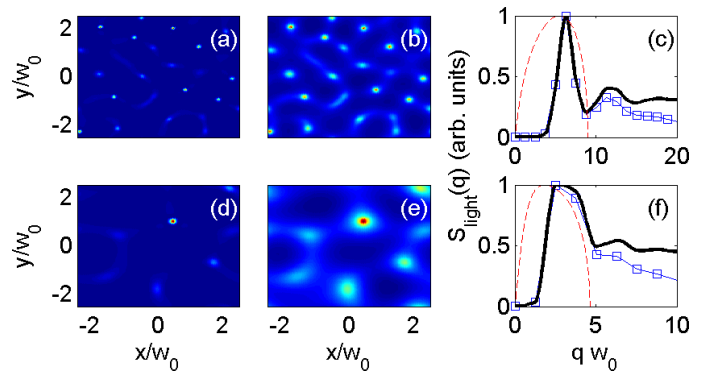}
\caption{(Color online) Filament pattern for two different 
fractal dimensions: (a) $D=1.3$ (distance $z=0.4z_R$); (d) $D=2.3$ (distance $z=z_R$);
(b,e) Corresponding index perturbations; 
(c,f) Corresponding $S_{light}$ (squares) averaged over $40$ noise realization; 
$S_L=S_{light}/I_F$ (solid line) 
and MI gain profile (dashed line)
($P=1000 P_r$, $\nu_p=10^{-4}$, $w_0=10^4 r_s$, $\xi=w_0$).
\label{filaments}}
\end{figure}

At the early stages of the evolution the input plane wave develops a spatial 
modulation, which spectrally corresponds to the MI gain bandwidth.  At longer propagation distances
narrow filaments are formed as evident by the snapshots (for a given noise realization) 
of the near-field intensity pattern $|a|^2$ in Fig. \ref{filaments}.
In the far-field such filaments are distributed along a circle due to the
axial symmetry of the problem (``conical emission''  \cite{centurion05}).
When the field propagate further, the filaments interact and form clusters (not shown) \cite{conti05PRE}. 
The fractal dimension $D$ not only affects the MI process, 
but also the number of generated filaments.
The global characterization of the filamentation process
requires a statistical description; we introduce the Fourier transform of the autocorrelation of $|a(x,y)|^2-\langle |a(x,y)|^2\rangle$
(the second term allows to get rid of plane-wave background), 
averaged over a large number of noise realizations and along curves $q=constant$ (due to radial symmetry).
This quantity, which is the equivalent for light of the material $S(q)$ and will be denoted
as $S_{light}(q)$, is shown in Fig. \ref{filaments}.
The peak positions of $S_{light}$ are determined by the average filament distance.
Additionally, so long as the filament positions and shape can be taken as independent
we can write  $S_{light}=S_L(q) I_F(q)$, where $I_F(q)$ is a {\it form factor}, which is 
approximately the average lineshape of the filaments, and $S_L(q)$ only depends on their statistical
distribution. These quantities are shown in Fig. \ref{filaments},
assuming for $I_F(q)$ a Gaussian profile (its width is the average filament waist).
The first peak in $S_{light}$, or $S_L$, corresponds to the
maximally amplified spatial harmonics of MI theory.

In conclusion we have developed a theoretical description
of laser beam filamentation in fractal aggregates and soft matter in general. 
The long-range statistical properties of multiple filaments reflect the material properties, 
and are strongly dependent by the fractal dimension of the aggregates.

We acknowledge support from the CNR-INFM Initiative for parallel computing, and
thank F. Sciortino (Univ. La Sapienza) for the useful discussions.

\end{document}